# Hydroxylated MgO (111) reconstructions: Why the case for clean surfaces doesn't hold water

J. Ciston[1*], A. Subramanian[1], L.D. Marks[1]

[1] Department of Materials Science and Engineering, Northwestern University

**Abstract**

We report an experimental and theoretical analysis of the $\sqrt{3} \times \sqrt{3}$-R30° and 2x2 reconstructions on the MgO (111) surface combining transmission electron microscopy, x-ray photoelectron spectroscopy, and reasonably accurate density functional calculations using the meta-GGA functional TPSS. The experimental data clearly shows that the surfaces contain significant coverages of hydroxyl terminations, even after UHV annealing, and as such cannot be the structures which have been previously reported. For the 2x2 surfaces a relatively simple structural framework is detailed which fits all the experimental and theoretical data. For the $\sqrt{3} \times \sqrt{3}$ there turn out to be two plausible structures and neither the experimental nor theoretical results can differentiate between the two within error. However, by examining the conditions under which the surface is formed we describe a kinetic route for the transformation between the different reconstructions that involves mobile hydroxyl groups and protons, and relatively immobile cations, which strongly suggests only one of the two $\sqrt{3} \times \sqrt{3}$ structures.

**Introduction**

Magnesium oxide is nearly ideal to serve as a model ionic material because it forms a very stable and simple rocksalt structure, and is free of any potentially challenging d-orbital transition metal bond hybridization. However, even this simple material presents a challenging problem when considering the formation of surfaces. When bulk MgO is truncated along the (111) direction, the alternating planes of $Mg^{2+}$ and $O^{2-}$ ions normal to the surface create an infinite surface dipole (and consequently infinite surface energy) in the strictly ionic formalism. However, as the true surface energy of a material can never be infinite, there must be some mechanism of valence compensation for nominally polar surfaces to mitigate the surface dipole. Valence compensation at a polar surface can be accomplished in three primary ways, all of which will be discussed in greater detail (see also [1,2] for an extensive review of polar oxide surfaces):

1. Redistribution of native species to create a surface stoichiometry which is valence neutral
2. Electronic compensation: creation of n-type or p-type states to metalize the surface and enable electron migration
3. Adsorption of foreign species in a manner that produces valence neutrality

The first compensation mechanism to restore valence neutrality is a change in surface stoichiometry usually achieved by the rearrangement of surface species and often results in terminations with partial vacancy coverage. To this end, Wolf proposed a 2x2 reconstruction of the (111) surface of rocksalt materials (2x2-oct) based upon valence-neutral octapolar units where ¾ of the top layer and ¼ of the second layer are composed of vacant sites [3,4]. This structure has been shown via density functional theory (DFT) surface energy calculations to be thermodynamically favorable over a wide range of oxygen chemical potentials [5,6]. While structures related to the 2x2 octapole have been experimentally suggested for NiO (111) [7], the octapolar structure has never been confirmed as an isolated phase on MgO (111). Finocchi *et al* interpreted their surface x-ray diffraction data in terms of the co-existence of an oxygen-terminated octapolar 2x2 reconstruction with a Mg-terminated α-phase with 2x2 periodicity [5]; a pure octapolar structure was not produced at the oxygen chemical potentials explored. Other surface structures with periodicities of √3x√3-R30° (Rt3), 2x2, 2√3x2√3-R30° have been observed on MgO (111) via transmission high energy electron diffraction (THEED) by Plass *et al* [8] and have been explained by models based upon cyclic oxygen trimer units. It is important to note that neither the 2x2 α-Mg structure nor the cyclic ozone 2x2 are stoichiometric and are therefore not inherently valence compensated reconstructions.

Recently, charge transfer at the MgO (111)-Rt3 surface was reported through the use of THEED and a structural model was proposed which combined the dipole-canceling effect of changes in surface stoichiometry with the creation of electron hole states near the surface [9]. This model utilizes the second method of valence compensation at the surface involving metallization and valence electron migration. However, since the band gap of MgO is very large (8.5eV), the creation of p-type states in the valence band requires a substantial amount of energy and would create a surface which is a stronger oxidant than molecular oxygen.

A third mechanism of valence compensation of polar surfaces is the adsorption of foreign species, such as water or hydrocarbons. This mechanism has the advantage of producing valence-compensated surfaces which do not require the energy intensive process of creating n-type or p-type intrinsic defects. It has long been known that surface hydroxyl groups are evident on MgO powders exposed to air and are persistent to over 500C in UHV [10-12]. It has also been shown both theoretically [13] and experimentally [14] that an OH terminated 1x1 structure on the (111) surface is very stable. The 1x1 surface of NiO (111) has also been shown experimentally to be OH-terminated [15]. This implies that it is improper to neglect the possibility of hydroxyl containing surface structures for samples annealed in or even exposed to air before introduction to a UHV environment as was the case in [5, 9]. To our knowledge, the *combination* of surface reconstruction to larger than 1x1 periodicities, and adsorption/dissociation of water has been largely neglected in the literature relating to MgO (111) surfaces.

In this paper, we report an experimental and theoretical analysis of the $\sqrt{3}\times\sqrt{3}$-R30° and 2x2 reconstructions on the MgO (111) surface combining transmission electron microscopy, x-ray photoelectron spectroscopy and reasonably accurate density functional calculations using the meta-GGA functional TPSS [16]. The experimental data clearly shows that the surfaces contain significant coverages of hydroxyl groups, even after UHV annealing, and as such cannot be the structures which have been previously reported. For the 2x2 surfaces a relatively simple structural framework is detailed which fits all the experimental and theoretical data presented herein. For the Rt3 there turn out to be two plausible structures, and none of the experimental or theoretical results can differentiate between the two within error. However, by examining the conditions under which the surface is formed we describe a kinetic route for the transformation between the different reconstructions that involves mobile hydroxyl groups and protons, and relatively immobile cations, which strongly suggests only one of the two Rt3 structures.

**Experimental Method**

MgO (111) single crystal substrates (MTI Corp., U.S.A., 99.9% pure, EPI polished 1 side) were cut into 3mm TEM size discs, mechanically dimpled, and thinned to electron transparency using a Gatan PIPS ion mill with 3-5 keV $Ar^+$ ions. It is important to note that the ion milling step tends to reduce the surface of oxides due to preferential sputtering of oxygen atoms. The samples were then annealed in a Carbolite STF 15/51/180 tube furnace between 800 - 1300C for 3-5 hours in dry oxygen or air to restore the nominal surface stoichiometry, recover the ion milling damage, and form a surface reconstruction.

To produce the Rt3 reconstruction while minimizing the effects of coarsening, a multi-step annealing process was required. The sample was first annealed at 800C in air for three hours to produce the 1x1 ordered surface and anneal out damage caused by the ion milling process.  The temperature was then raised to 1000C and held for three hours to obtain the Rt3 reconstruction.  A simpler 1-step anneal at 1000C for 3 hours also produces the Rt3 reconstruction, though not as reliably as the multi-step process, indicating sensitivity to damage from the sample preparation.  This also suggests that the initial formation of the ordered 1x1H surface (which forms at 800C) is required for the Rt3 reconstruction to form.  If the anneal is performed in dry oxygen, the Rt3 reconstruction does not appear leaving only a 1x1 ordered surface.  To produce the 2x2 reconstruction, the sample is annealed in dry $O_2$ at 1300C for three hours.  At this higher temperature, an 800C pre-annealing step is unnecessary presumably due to the large increase in surface diffusion.

Electron diffraction experiments were performed with both Hitachi H-8100 and Hitachi UHV-H9000 transmission electron microscopes (TEM) at 200kV and 300kV respectively using off-zone-axis electron diffraction patterns. Conventional bright-field and dark-field imaging was also used but these results are not reported here. Diffraction data was recorded on film negatives with exposure times ranging from 2-90s and digitized to 8 bit precision and 25μm pixel pitch using an Optronix P-1000 microdensitometer calibrated to be linear. Intensities were measured using a cross-correlation techque in EDM software [17, 18].  For the Rt3 structure, a total of 527 measurements which were reduced to 19 p3m1 symmetry unique reflections (p6mm Patterson symmetry) using a Tukey-biweight method to a resolution of 2 $nm^{-1}$, and similarly for the 2x2 structure 703 measurements were reduced to 27 p3m1 symmetry unique reflections to 2$nm^{-1}$.  Typical diffraction patterns of the Rt3, and 2x2 reconstructions are shown in Figure 1.

Annealed samples were then transferred to a UHV chamber (base pressure ~$1x10^{-10}$ torr) equipped with an x-ray photoemission spectroscopy (XPS) source and hemispherical analyzer, a UHV in-situ resistive heating stage, and a Hitachi UHV-H9000 300kV TEM.  Thus chemical information from XPS can be correlated with structural information from diffraction without exposure to air (or water vapor).  XPS data was collected with an Al kα source and an analyzer takeoff angle of 60 degrees with respect to the surface normal in an effort to maximize surface sensitivity; linear background subtraction and symmetric Gaussian functions were used for peak fitting.  To correct for sample charging, the scan energies were offset such that the C-1s peak had a value of 284.5eV for all scans.  To investigate the possible presence of surface contaminants, a wide-band XPS scan was collected in the 1400eV→0 binding energy range, and the only features not related to magnesium or oxygen were a small C-1s peak and an occasional trace of an argon 2p peak from implantation in the ion milling step.   Fine scans of the O-1s region at 0.1eV resolution were collected to investigate the possibility of a hydroxylated surface on the Rt3 and 2x2 reconstructions.  XPS data for the Rt3 samples was collected using a low-background sample holder, but data for the 2x2 surface was collected using a Mo sample support which introduces an artifact peak in the O-1s and C-1s regions offset 4-5 eV lower in binding energy from the primary signal due to differences in charging between the insulating sample and the grounded Mo holder.  This offset is sufficiently large not to interfere with the high binding energy features of interest for this experiment.

**Theoretical Method and Calibration**

All DFT calculations were performed using the all-electron Wien2k APW+lo code [19]. To be consistent, we used muffin tin radii of 0.6, 1.2, and 1.63 for H, O, and Mg for all the (111) surface structures, and RKMAX/min(RMT) of 2.75/0.6 in all cases. Conventional slab-models were used, with the distance between the surfaces 8-10 Angstroms which numerical tests indicated to be large enough. The number of k-points was adjusted to a density of approximately 49 points per 1x1 Brillouin-Zone in the reciprocal-lattice plane normal to the surface. The MgO surfaces were calculated spin-unpolarized; tests indicated that including spin polarization had no effect. All proposed structures were geometrically relaxed such that all residual forces were less than 0.1eV/Å. The surface energies were determined by subtracting the energy of the appropriate number of bulk unit cells, the latter calculated in a hexagonal supercell with axes [1-10]x[10-1]x[333] and a comparable density of reciprocal-space sampling points. Because of symmetry requirements the sizes of the cells differed for the 1x1, Rt3, and 2x2 structures, and the number of layers in each slab is given later when the results are presented.

For the calibration structures (see below) the isolated molecules were calculated in cells of between 15 and 20 au size; this was large enough to avoid interactions. Except for the $O_2$ molecules where a slightly smaller RMT of 1.1 was used, the same RMTs as for the (111) surfaces were used. For the MgO (001) surface slightly larger RMTs of 1.9 and 1.8 for Mg and O were used with an RKMAX of 7.25 with 13,15,17 and 19 layer slabs to check the convergence versus slab size. As would be expected, the surface energy for the smaller 13 layer slab was well converged.

What functional one should use is a subtle issue. While DFT calculations of surface structures and energies have become common, unfortunately there are some well-known fundamental problems which raise questions about their validity and accuracy. For an oxide system the calculational method must be able to adequately represent:
  a) The ionicity, both in the bulk and how it changes at the surface.
  b) The covalency, both in the bulk and how it changes at the surface.
  c) The long-range energy of the electron decay into vacuum, in effect the jellium surface energy contributions.
  d) The energetics of any chemisorbed molecules, both in the gas-phase as well as at the surface.

For MgO the problems are not overwhelming. The ionicity and covalency are fairly well represented by conventional LDA [20] or the conventional generalized gradient approximation (GGA) as defined by the PBE functional [21], with a GGA in general being better, although neither are unconditionally accurate. It is well-established [22-24] that LDA does better for the jellium contribution but this is due to a fortuitous cancellation. At present the most accurate available method for the long-range jellium contributions is to use the exchange-correlation energy from the meta-GGA TPSS functional [16]. While it is not currently feasible to implement this in a completely self-consistent fashion within a conventional DFT code, it is known that a first-order approximation when only the exchange-correlation energy terms from the TPSS functional are included works quite well. To verify this, some representative numbers for small-molecule test cases are shown in Table 1.

For the chemisorbed molecules relevant herein the problems are also easily solvable. A LDA approach badly overestimates the bonding energies; PBE-GGA performs better but is still inadequate for double-bonds such as in the $O_2$ molecule whereas the TPSS meta-GGA does rather well in all cases (see Table 1) although it is still far from perfect. To estimate the error of the DFT surface energies, we note that for the small molecules, the error in the TPSS calculations is slightly smaller than the difference between PBE and TPSS. Additionally, the PBE errors are approximately twice as large as the TPSS errors on average. Using the values for the different surface structures shown later, we estimate the error in the DFT calculated surface energies to be <TPSS-PBE> (0.05eV) per 1x1 surface unit cell for the TPSS functional and approximately 2*<TPSS-PBE> (0.1eV) for the PBE functional.

**Results**

**Direct Methods**

In this section we present structures solved from direct methods analysis of THEED data for both the Rt3 and 2x2 reconstructed samples. Direct methods are a set of routines that utilize statistical relationships and self-consistent equations relating diffracted beam phases to determine possible scattering potentials with measured diffraction amplitudes and symmetry as inputs [25, 26]. It is important to note that because direct methods require no *a–priori* information about the structure, the proposed scattering potentials may or may not represent structures which are physically and chemically possible. It is also possible that multiple unrelated structures may yield an equally good fit to the experimental amplitudes as both the data and structures themselves form non-convex sets [27]. As a consequence of the non-convexity of the problem, it is difficult to determine the statistical significance between refinements of two fundamentally different structures in the same way that one must not use a Hamilton R-test [28] to compare the fit of two fundamentally different functions to a given dataset. Additionally, because all surface reflections which overlap those of the bulk are discarded, the direct methods output does not provide any information about the bulk registration of the proposed structures which must be determined by other means. One must also use care in ascribing particular atom species to peaks in the direct methods potential, especially in the case of MgO where the scattering difference between species is not large, and by adjusting the Debye-Waller terms for the atoms they can be made to be essentially identical.

The dataset used for the solution of the Rt3 reconstruction has already been published elsewhere [9]. The structure proposed in that analysis comprised a Mg-terminated surface with 2/3 of the bulk-like Mg atoms missing (Rt3-Mg). Although the surface is Mg-terminated, a proper surface excess sum shows that this structure is oxygen-rich (1/6 O per 1x1 unit cell) which is consistent with the findings of [29]. A key point missed in the prior analysis of this data was the existence of another related solution from direct methods phase restoration that implies adding 2/3 monolayer of oxygen to the bulk termination (Rt3-O) rather than subtracting 2/3 monolayer of Mg, again yielding an O-rich non-stoichiometric surface. Indeed the Rt3-O structure is the Babinet of the Rt3-Mg structure, i.e. the amplitudes of all reflections except the 1x1 family are the negative of those for the Rt3-Mg structure and by definition the two are indistinguishable from one another using in-plane electron diffraction data. (It should be mentioned that Babinet solutions are a well-known [30], but sometimes forgotten problem with direct methods for surfaces.) Two additional structures are possible, formed by the addition of a hydrogen atom per unit cell to the Rt3-Mg (in the second layer) or Rt3-O (as an OH termination) to produce Rt3-OH and Rt3-MgH structures. While with care hydrogen can be detected at a surface in a diffraction experiment once other bonding terms are properly accounted for [31], in general this is very difficult and thus not discernable from examination of the direct methods potentials. Direct methods potential maps for the two Rt3 solutions are shown in Figure 2 with overlays of the proposed structures. The Mg and O positions of the un-hydrogenated Rt3 structures are essentially identical to the hydrogenated structures and therefore not shown.

Direct methods solutions of the 2x2 structure yields a potential shown in Figure 3 with features in an open hexagonal arrangement. This potential immediately suggests the structure is not the proposed octapole structures of Wolf [3]. The cyclic ozone trimer structures of Plass *et al* [8] are also inconsistent with the direct methods results presented here, however Plass utilized reducing UHV annealing conditions rather than the oxidizing anneals presented herein. The THEED data can be explained by an epitaxial α-phase structure (Figure 9), similar to Finocchi *et al* [5] (2x2-α), that retains the underlying cation framework (with 1x1 periodicity, and as such is absent from the direct methods surface potential map) and contains three symmetry inequivalent surface sites which may be populated with Mg, O, or other species to account for the features in the potential map (Figure 3). For completeness, we note that in this particular case reversing the sign of all reflections except the 1x1 (Babinet operation) does not change this structure.

It is a subtle point that all of the atomic sites in the potential maps for the 2x2-α framework and the two Babinet Rt3 solutions are high-symmetry special sites wherein the x,y positions of the atoms are fixed (in this case 0,0; 1/3,2/3; and 2/3,1/3). Therefore, any attempt to refine these structures will only probe temperature factors, subtle changes in underlying bulk positions, and bonding effects as the primary degrees of freedom. Because the atomic scattering factors of Mg and O atoms are similar, one may populate these special sites in a variety of ways and use temperature factors, bonding effects, or add hydrogen atoms or holes to compensate for differences in the refinement to produce R-factors of 0.1-0.2. As a result, all permutations of site populations with Mg, O, or OH species can be refined to fit equally well which necessitates the use of complementary techniques to confirm the validity of the refinements. In a most fundamental sense, the electron diffraction data for these reconstructions may only be used to exclude incompatible structures, but not determine unequivocally which of the structures we have proposed is correct, only that they are consistent with the direct methods potentials. Therefore, one needs to perform DFT structural relaxations to obtain out-of-plane positions, estimate surface energies, and verify the stability of the proposed structures. As mentioned previously, surface diffraction techniques are limited by the relative insensitivity to adsorbed hydrogen, for which we instead turn to XPS data in the following section.

**XPS**

Lazarov *et al* [14] have demonstrated the presence of OH bonds at the MgO (111)-1x1 surface and correlated this finding with a shoulder ~2eV on the high binding energy side of the O-1s XPS peak. We have also investigated the detailed behavior of the O-1s peak as a function of UHV annealing temperature to obtain information about non-native adsorbates.

XPS scans of two independent ex-situ annealed Rt3 TEM specimens (Rt3-A and Rt3-B) both exhibited a shoulder on the O-1s peak located 1.7±0.14 eV (average) higher in binding energy than the primary peak (Figure 4).  See Table 2 for the complete O-1s XPS results.  This shoulder position is consistent with the presence of O-H bonds at the surface.  The sample Rt3-A was then annealed at 400C for 12 hours in-situ in an attempt to desorb molecular water and residual volatile organic carbon.  Post-anneal XPS shows no change in either the area or position of the hydroxide shoulder, though the C-1s peak does decrease in area.  This is consistent with the disappearance of the HOH bending mode in infrared spectroscopy studies above 100C [10-12], and implies that the remaining OH stretching interaction, which persists to 500C in UHV, is due to surface hydroxyl groups and not physisorbed water.  Sample Rt3-B was annealed at 750C for 3 hours in-situ after which the O-1s shoulder nearly disappeared and the primary O-1s peak shifted 0.8eV higher in binding energy while maintaining the same width.  Sample Rt3-B was then transferred to the UHV TEM without breaking vacuum, where electron diffraction exhibited no evidence of the Rt3 reconstruction.  This suggests that chemisorbed OH groups are an integral feature of the Rt3 structure.

XPS scans of an ex-situ annealed 2x2 TEM specimen also shows a high energy shoulder on the O-1s peak at 1.8±0.07 eV with a shoulder area 1.5x as large as that observed on the Rt3 specimens suggesting relatively larger surface coverage of OH groups (Figure 5 and Table 2).  As with the Rt3 sample, annealing to 350C in UHV does not significantly affect the XPS spectrum suggesting that the shoulder feature of interest is not due to physisorbed water.  As previously mentioned, the XPS data for the 2x2 sample was not collected using the same low background holder used for the Rt3 samples and the additional feature ~4eV lower in binding energy is due to signal from the grounded Mo ring which experiences a different amount of charging than the insulating sample.  This low energy peak grows as the sample is misaligned, moving more of the holder into the visible electon analyzer window.  The C1-s peak exhibits an identical 4eV peak splitting confirming this interpretation.  Further analysis of the C-1s peak shows no evidence of a high energy shoulder to suggest the presence of C-O bonds.  This supports our attribution of the O-1s shoulder to OH bonding and confirms that while the surface is not 100% clean, the carbon is present in a physisorbed state.

**DFT Energetics at 0 K**

Surface energies for a wide variety of proposed structures were calculated and normalized to the 1x1 unit cell area using both the PBE-GGA and the TPSS functional as described above.  Figure 6 is a compositional summary of all of the structures plotted on a surface-excess ternary phase diagram in units of excess atoms per 1x1 cell.  Structures to the left or right of the vertical line are not valence-balanced and would thus require a valence compensation mechanism involving the creation of either n-type or p-type states at the surface.  However, the band gap in MgO is so large that the creation of holes in the valence band (O-rich composition) would make the surface more oxidizing than molecular oxygen.  Similarly, adding electrons to the conduction band (Mg-rich structure) results in a highly reducing surface. We also calculated the energies of $H_2O$ and $O_2$ molecules using both functionals since these are needed to assess the thermodynamics. The full set of calculated surface energies can be found in Table 3.  Refined positions for these different models are available in the online EPAPAS repository in the format of conventional crystallographic cif files.

**Hydroxylated 1x1 Structure**

Of the configurations sampled, we found the fully hydroxylated 1x1 surface to be a very stable structure.  This agrees with the experimental and theoretical analysis of this reconstruction by Lazarov *et al* [14] and the relaxed structure is shown in Figure 7.

**Rt3 Structures**

Previous studies of the Rt3 reconstruction under very similar annealing conditions [9] have suggested a magnesium-terminated surface structure with 2/3 of the bulk-like Mg atoms absent in the surface layer (Rt3-Mg) [9].  As mentioned earlier, the direct methods result from electron diffraction cannot distinguish between Mg and O terminated surfaces due to their similar scattering cross-sections.   The XPS data suggests that there are hydroxyl groups at the surface, which allows us to discard the Rt3-O and Rt3-Mg structures in favor of their hydroxylated equivalents.  However, since the coverage for the Rt3-MgH (Figure 8a) and Rt3-OH (Figure 8b) structures are the same neither XPS nor direct methods can unambiguously determine the correct configuration. In the DFT calculations the structure terminated with one oxygen hydroxylated per unit cell (Rt3-OH) was found to be 0.1eV/1x1 lower in energy than the Mg-terminated structure with a second-layer hydroxide (Rt3-MgH).  While this energy difference is small, the energies are two sigma apart with respect to the approximated DFT error (0.05eV for TPSS) yielding a >90% confidence in the Rt3-OH structure.  However, as this error value is an estimate, a more conservative error estimate of 0.1eV is on the same order as the different in DFT surface energy placing a lower bound on the confidence level at one-sigma (67%) and is somewhat precarious.

**2x2 Structures**

Turning to the 2x2 reconstruction, there are several structures with the configuration matching the direct methods results that are stable when undergoing DFT relaxation. The primary difference between these structures is the selection of atoms to fill the three surface sites. Figure 9 is a schematic diagram of the generic 2x2-α structure where sites 1, 2, and 3 can be filled with either Mg, O, OH, or $H_2O$ groups. Table 3 contains the PBE and TPSS surface energies for the most stable configurations, and also includes the Mg- or O-terminated octapolar structures for comparison. We note that by allowing one $H_2O$ molecule per unit cell to dissociate one may create stable hydroxylated 2x2 octapolar structures (O-Oct → OH-Oct and Mg-Oct → MgH-Oct) that at zero water chemical potential are lower in energy than either dry octapole and are lower in energy than all of the 2x2-α-OH structures at all water chemical potentials. However, none of the octapolar structures (neither dry nor hydroxylated) are consistent with the electron diffraction data presented herein.

One set of 2x2-α structures has all three surface sites populated by oxygen with 2 of the 3 oxygens per 2x2 unit cell terminated with hydrogen (2x2-α-OH). There are two structures of this configuration which are indistinguishable in the DFT energetics (and 0.3eV/1x1 lower in energy than the third permutation). It is important to note that all of the 2x2-α-OH structures contain 1/4 $H_2O$ per 1x1 cell, whereas the Rt3-OH and Rt3-MgH contain 1/6 $H_2O$ per 1x1 cell which is consistent with the aforementioned XPS finding that the 2x2 samples exhibited approximately 1.5x the OH coverage of the Rt3 samples. The two lower energy 2x2-α-OH1,2 structures utilize the uni-coordinated oxygen surface site (site 3 in Figure 9), whereas the third (higher energy) α-OH3 arrangement shuns this site in favor of occupying both surface oxygen tetrahedral sites (sites 1 and 2 in Figure 9). A DFT-relaxed structural model of the 2x1-α-OH2 surface is shown in Figure 10.

A related 2x2-α structure has only 2 of the 3 surface sites occupied by oxygen (2x2-α-O), leaving the third vacant. This is a stoichiometric MgO termination, and is higher in energy than either the Mg- or O- terminated octapole structures with which it can be directly compared. However, since it has no hydroxide it is inconsistent with the XPS data. For completeness, we note that it is possible to fit the electron diffraction data with a combination of the α-O structure and the related hydroxylated ones. Note also that the 1x1H, Rt3-OH, α-O, and α-OH structures all lie on the same water desorbtion line in the surface excess phase diagram in Figure 6 and are therefore compositionally related to one another according to the surface coverage of $H_2O$.

The last possibility is a Mg-terminated 2x2-α-Mg structure [5]. However, this requires a highly reducing environment to be stable which is completely inconsistent with the experimental conditions so this structure can be unconditionally ruled out.

**Thermodynamics**

There are several ways to interpret the DFT data. A conventional method would be to plot the energies versus a relevant chemical potential, for instance that of $H_2O$ and $O_2$ taking the chemical potential of MgO as a reference. Recently, such a surface phase diagram of MgO (111) has been calculated for a variety of non-hydrous MgO (111) structures with respect to oxygen chemical potential [6]; we show a similar plot in Figure 11 for the water case. This assumes that the surface is in equilibrium with the gas, which may not be the case. We note that even with respect to oxygen we should not assume global equilibrium, rather only local equilibrium similar to the $TiO_2$ (001) surface [32].

An alternative is to consider the energetics for a fixed surface excess of (for instance) $H_2O$, and consider separately how this surface excess changes with sample treatment. The stable phases for the valence-compensated structures can then be generated using a conventional convex-hull construction, connecting all points on an energy-composition diagram and taking the lowest energy combination. This is shown as a tie line in Figure 12.

If we assume full equilibrium with gas-phase water, then from Figure 11 the hydroxylated 1x1 is stable above a water chemical potential of -2.2±0.1 eV, and at lower (more negative) chemical potentials there should be a mixture of the Mg- and O-terminated octapoles. It should be noted that the hydroxylated octapoles are never the lowest energy structures for the values of $H_2O$ chemical potential considered here. Alternatively, if we consider the surface concentration of water as fixed one expects a two-phase co-existence of the hydroxylated 1x1 and octapoles as indicated in Figure 12. This prediction does not agree with the experimental data and implies that the observed structures are kinetically metastable.

**Kinetic Model**

Since the proposed 2x2-α-OH structure cannot be explained by the DFT thermodynamics, either the functionals are not yet good enough or we need to consider kinetics. Additionally, the nearly degenerate surface energies of the Rt3-OH and Rt3-MgH structures prompts one to consider which of the structures is more easily accessible. From our analysis of comparable small molecules it is unlikely (but not impossible) that the DFT energies are too inaccurate. As we will show here a relatively simple kinetic model can explain the data.

We first consider the potential reordering process as water desorbs starting from a 1x1H structure. The formation of the Rt3-MgH structure from a 1x1H precursor requires quite a substantial reordering, a process by which the surface must transition from OH termination to Mg termination while retaining 1/3 of the initial hydrogen adsorbates in what is now the second layer of the structure. This can be accomplished either by partial desorbtion of hydrogen coupled by diffusion of Mg atoms to the surface, or the removal of all surface OH groups from the 1x1H and 2/3 of the next magnesium layer followed by re-adsorbtion of hydrogen to 1/3 of the second layer oxygen sites. Both of these processes require a nett desorbtion of one $H_2O$ molecule per Rt3 unit cell in addition to significant bulk exchange of cations.

The formation of the Rt3-OH structure may follow a much more direct pathway as it retains the cation framework of the 1x1H structure and requires only a single proton per Rt3 cell to hop to an adjacent site causing re-association and molecular desorbtion of $1/3 H_2O$ from each 1x1H cell; see Figure 13a for a diagram of the 1x1H→Rt3-OH transition. Therefore the formation of the Rt3-O structure from a 1x1H precursor can be described solely by the desorbtion (and re-association) of $H_2O$ molecules from the surface. By calculating the chemical potential of water vapor for various temperatures and pressures,

$$\mu(T,P) = \mu_o - (T-T_o)S(T) + RT \ln\left(\frac{P}{P_o}\right) \qquad \text{Equation 1}$$

where values with the subscript $_o$ refer to STP conditions with $\mu_o$ and $S(T)$ are taken from [33, 34], we can determine at what conditions the driving force for water desorption makes the 1x1H→ Rt3-OH transition thermodynamically favorable. Assuming an air anneal with 50% relative humidity (~0.02 atm water), the Rt3-OH structure becomes thermodynamically favorable above 200C (2.9eV/$H_2O$). This transition is not observed experimentally below 950-1000C and the Rt3-OH structure is air stable for months at room temperature, both of which suggest that the kinetics of the transition are relatively slow.

The activation barrier for the water desporbtion event may be estimated given the experimental values of annealing temperature and time. From the standard kinetic rate equation,

$$\gamma = f \exp\left(\frac{-\Delta E}{k_B T}\right) \qquad \text{Equation 2}$$

where $\gamma$ is the turnover rate, f is a frequency prefactor often referred to as an attempt frequency, $\Delta E$ is the activation barrier, $k_b$ Boltzmann's constant, and T the absolute temperature. For the 1x1H → Rt3-OH transition, one water molecule must desorb per Rt3 unit cell (27Å$^2$) for the entire area of the 3mm disc (2.8x10$^{15}$ Å$^2$) in the 10800s annealing time implying a turnover rate of 9.7x10$^{-9}$ per second. At an annealing temperature of 1273K, the attempt frequency for the desorption of water from the MgO (100) surface is approximately 9x10$^{14}$ attempts per second [35]. While the attempt frequency is derived from a different surface of MgO than we are studying here, the activation energy is logarithmically dependant upon this value, so this estimate should be sufficient. Solving equation 2 for the activation energy

$$\Delta E = -k_B T \ln\left(\frac{\gamma}{f}\right) \qquad \text{Equation 3}$$

yields an estimate of 1.3 eV for water desorption from the 1x1H surface. While it is not known if the full three hours of annealing is necessary to produce the Rt3-OH reconstruction, the activation barrier is weakly dependant upon annealing time. It is important to note that this estimate includes the barrier necessary for hydrogen migration to an adjacent site to facilitate $H_2O$ recombination before the desorption event.

Turning to the transition from the Rt3-OH structure to the proposed 2x2-α-OH structure, it is again instructive to investigate potential kinetic pathways in order to determine why the 2x2-oct structure is not achieved despite its lower energy over the calculated range of water chemical potential. The removal of ½$H_2O$ from each Rt3-OH unit cell produces a structure (α-O) with a 2x2 unit cell that is OH-free, retains the cation framework, fits the direct methods potential from the 2x2 diffraction data, and is valence neutral; see Figure 13b for a diagram of the Rt3-OH→2x2-α-O3 transition. It should be noted that this transition requires the diffusion of a proton to a second nearest neighbor oxygen site and as such is predicted to have a larger activation barrier than the 1x1H→Rt3-OH transition. Again, by calculating the chemical potential of water vapor we can estimate the temperature required at 50% relative humidity to make the Rt3-OH→2x2-α-O3 transition thermodynamically favorable is 700C (4.5eV/$H_2O$). Again, this is several hundred degrees lower than the annealing temperature at which the 2x2 structure is observed experimentally suggesting rather slow kinetics.

As the temperature falls below 125C (2.7 eV/$H_2O$) on the cooling side of the annealing process, the dissociation of one water molecule per 2x2 cell becomes energetically favorable, forming the proposed 2x2-α-OH1, α-OH2, or some combination of the two structures. Before dissociating to form a 2x2-α-OH structure, water molecules must first adsorb to the surface; however, DFT structural relaxations suggest that adsorbed molecular water is unstable and dissociates to the 2x2-α-OH configuration even at 0K. This is consistent with our XPS results which exhibited no change in the O-1s shoulder after UHV heating up to 500C implying that any O-H bonding is due to hydroxyl groups, and not molecular water. The 2x2-α-O3 structure is the lowest energy stoichiometric α-type structure at all water and oxygen chemical potentials and contains a vacancy at surface site 3. In a DFT relaxation, molecular water adsorbing to site 3 dissociates to form the 2x2-α-OH1,2 structures depending upon the orientation of the hydrogen atoms with respect to surface sites 1 and 2. As noted before, the two lower energy α-OH1,2 configurations both contain a hydroxyl group at site 3 which also suggests the initial molecular adsorption of water to site 3 from a kinetic standpoint.

Chemical potential analysis also suggests that water should desorb from the 2x2-α-OH structures at temperatures above -45C at $H_2O$ partial pressures below $10^{-8}$ torr, but XPS results presented previously show that the 2x2 surface is OH-rich until at least 350C in UHV conditions. This again suggests a rather large activation barrier for the desorption of water from the MgO (111) surface.

We have presented a structural evolution model for the 1x1H→Rt3-OH→2x2-O→2x2-OH system which retains the cation framework and is driven by ad/desorption and dissociation of water molecules at the MgO (111) surface (Figure 12). The overall agreement between the Rt3 and 2x2 structures obtained by direct methods and DFT surface energetics is quite good. When plotted on a ternary surface excess phase diagram, these five structures all lie on a water desorption curve connecting 2x2-O to $H_2O$ consistent with the transition model (Fig 6).

**Discussion**

The combination of electron diffraction and direct methods has made great strides in the last decade toward the goal of "easy" solution of surface structures.  However, inherent limitations such the inability to measure surface spots coincident with bulk reflections, dynamical scattering, non-convexity of the diffraction data and structural models, and the possibility of Babinet solutions means that while some structures can be unconditionally ruled out (such as the 2x2-octapole here), it is sometimes the case that more than one solution is equally likely (as in the Rt3 example).  In such cases, one must bring additional information to bear upon the problem either by probing chemistry (XPS) or other properties.  One may also attempt to compare calculated surface energies to sort out the thermodynamics, but while there is every reason to believe that the DFT energies reported here using the TPSS functional are superior to traditional PBE-GGA methods, it is imperative that one considers the error bars placed on the surface energy calculations (0.05-0.1eV per 1x1 cell in this instance).  In the case of Rt3-OH vs. Rt3-MgH structures, the difference in DFT surface energy is just on the edge of being significant based upon the estimated errors and as such it is not possible to say that the DFT calculations prove that the Rt3-OH structure is unconditionally correct, only that it is more likely than the Rt3-MgH structure within DFT error for the functionals and parameters utilized herein.

It is also useful to explore the possible kinetic pathways of structural transition, which in this case strongly suggests that the Rt3-OH structure should readily form given a 1x1H precursor.  The kinetic model is also consistent with the experimental observation that while the 2x2-octapole structures are energetically favorable, they may be kinetically inaccessible.  It does not follow that the octapolar structures will never be obtained under any annealing conditions, however, the kinetics of oxygen and to a greater extent magnesium exchange that are required for the extensive rearrangement of atoms to transition from the 2x2-α phase to an octapolar structure are likely to be far slower than that of hydrogen hopping which implies that the water-desorbtion driven structural transitions proposed herein will occur more rapidly.  Higher temperatures and longer times in dry annealing conditions will generally be required to allow for the relatively sluggish cation diffusion processes to produce the 2x2 octapole structures.

While the proposed transition model in Figures 12-13 is consistent with XPS, THEED, and DFT observations, it is only the endpoints (1x1-H, Rt3-OH, 2x2-OH) which have been conclusively observed experimentally.  The proposed kinetic model behaves something like the interaction of a hydroxide gas with the MgO (111) surface over a fixed-cation framework.  As such, there may be more than one stable conformation of the hydroxide species for each reconstructed periodicity (as in the 2x2-α-OH1,2 case).  Therefore, we are currently investigating the kinetics computationally using the climbing-image nudged elastic band method[36, 37] to determine activation barriers and multiple local minima for each proposed structure [38].  An initial estimate of the activation barrier for hydrogen atoms to hop to an adjacent oxygen site is about 0.8 eV, which the application of equation 3 implies that hydrogen exchange will occur at 100% of sites within 5 minutes of annealing at 1000 C.  Confirmation of the proposed transition pathway may also be obtained experimentally through the use of an environmental TEM with an in-situ heating stage which would enable the collection of THEED data while varying the $O_2$ and $H_2O$ chemical potential.  Activation energies for the proposed desorbtion events may be probed by performing Temperature Programmed Desorbtion (TPD) studies and observing rate constants.  Both of these experiments are beyond the scope of this paper, but will be the subject of future studies.

The dominant role which water plays in the formation of reconstructions on the MgO (111) surface as a result of the high surface mobility of hydrogen atoms is suggestive of the possibility of water-driven reconstructions on other polar oxide surfaces - even for experiments performed in UHV since OH groups may persist to several hundred degrees Celsius.  Due to the weak scattering behavior of hydrogen at a crystal surface, the difference between oxygen-terminated and hydroxyl-terminated polar surfaces is nearly indistinguishable by diffraction methods alone and may have been missed in prior studies.  We are also presently extending the methodology presented herein to the polar NiO (111) surface which is nearly identical to MgO (111) from a structural point of view.  However, the highly correlated d-electrons in Ni pose well-known computational difficulties not encountered in the MgO system, such as too much hybridization between the metal 3d and oxygen 2sp states which reduces the ionicity of the bond.  Initial results suggest that the proposed surface phase transition pathway determined for the case of MgO are also correct for NiO and may be general for the (111) surface of rocksalt metal oxides [39].

**Acknowledgements**

This work was supported by the National Science Foundation DMR-0455371 (JWC and LDM) and DMR-0075834 (AKS). The Hitachi-H8100 TEM work was performed in the EPIC facility at the NUANCE Center at Northwestern University. NUANCE Center is supported by NSF-NSEC, NSF-MRSEC, Keck Foundation, the State of Illinois, and Northwestern University.

**Tables**

|              | PBE    | TPSS   | Expt   | \|PBE Err\| | \|TPSS Err\| | \|PBE-TPSS\| |
|--------------|--------|--------|--------|-------------|--------------|--------------|
| $O_2$        | 6.096  | 5.493  | 5.213  | 0.882       | 0.279        | 0.603        |
| MgO (001)    | 0.515  | 0.577  | 0.650  | 0.135       | 0.073        | 0.061        |
| $H_2O$       | 10.091 | 10.210 | 10.053 | 0.038       | 0.157        | 0.119        |
| MgOH         | 8.005  | 8.082  | 7.569  | 0.437       | 0.513        | 0.077        |
| MgO          | 2.973  | 2.741  | 2.572  | 0.401       | 0.169        | 0.232        |
| $O_3$        | 8.055  | 6.927  | 6.413  | 1.642       | 0.514        | 1.128        |
| $H_2O_2$ | 12.143 | 11.952 | 11.617 | 0.526 | 0.334 | 0.192 |
| OH | 4.749 | 4.776 | 4.640 | 0.110 | 0.137 | 0.027 |
| $H_2$ | 4.497 | 4.837 | 4.743 | 0.246 | 0.095 | 0.340 |
|  |  |  | **Averages** | **0.49** | **0.25** | **0.31** |

Table 1: Small molecule atomization energies (corrected for the zero-point energy) as well as the surface energy of MgO (001) calculated with PBE and TPSS functionals compared with experimental data. All energy values are in units of eV. The atomization energies were taken from the CCODB database [40] and the surface energy for MgO (001) from [41] which may be a slight overestimate since it is measured as a zero-velocity limit for cleavage.

|  | O-1s Peak | Shoulder Shift (eV) | Shift Error | Relative Area | Area Error |
|---|---|---|---|---|---|
| Rt3-A, 30C | 529.9 | 1.8 | 0.25 | 0.28 | 0.05 |
| Rt3-B, 30C | 529.8 | 1.6 | 0.13 | 0.20 | 0.02 |
| Rt3-B, 400C | 529.8 | 1.7 | 0.28 | 0.22 | 0.06 |
| Rt3-A, 700C | 530.7 | 1.7 | 0.43 | 0.11 | 0.03 |
| 2x2, 30C | 529.8 | 1.8 | 0.07 | 0.38 | 0.03 |
| 2x2, 350C | 529.7 | 1.7 | 0.10 | 0.45 | 0.03 |

Table 2: Results of XPS O-1s scans for various Rt3 and 2x2 samples annealed in UHV at the indicated temperature.

**1x1 Structure**

|  | PBE (eV) | TPSS (eV) | Layers |
|---|---|---|---|
| 1x1-H | 0.11 | 0.00 | 15 |

**Rt3 Structures**

|  | PBE (eV) | TPSS (eV) | Layers |
|---|---|---|---|
| Rt3-Mg | 0.94 | 0.96 | 13 |
| Rt3-OH | 0.94 | 0.97 | 13 |
| Rt3-MgH | 1.10 | 1.06 | 13 |

**2x2-α Structures**

|  | Site1 | Site2 | Site3 | PBE (eV) | TPSS (eV) | Layers |
|---|---|---|---|---|---|---|
| α-Mg | Mg | Mg | Mg | 3.45 | 3.28 | 13 |
| α-O1 | N/A | O | O | 2.16 | 2.18 | 11 |
| α-O2 | O | N/A | O | 2.42 | 2.46 | 11 |
| α-O3 | O | O | N/A | 1.73 | 1.73 | 11 |
| α-OH1 | OH | O | OH | 1.03 | 1.06 | 13 |
| α-OH2 | O | OH | OH | 1.01 | 1.05 | 13 |
| α-OH3 | OH | OH | O | 1.29 | 1.33 | 13 |

**2x2 Oct Structures**

|  | PBE (eV) | TPSS (eV) | Layers |
|---|---|---|---|
| O-Oct | 1.19 | 1.07 | 19 |
| Mg-Oct | 1.27 | 1.14 | 19 |
| OH-Oct | 0.65 | 0.61 | 19 |
| MgH-Oct | 0.86 | 0.81 | 19 |

Table 3: DFT calculated surface energies for various MgO-(111) structures referenced to 1x1H-TPSS. Number of layers (not including hydrogen) given for each slab calculation.

**Figure Captions**

FIG. 1. Electron diffraction patterns taken from MgO samples with a) $\sqrt{3}\times\sqrt{3}$-R30$^o$ and b) 2x2 reconstructions. The surface unit cell has been outlined for each image. Images are composites of multiple exposures to increase dynamic range.

FIG. 2. (color online) Rt3 direct methods potential maps with unit cell outlined overlaid with models of the a) Rt3-MgH structure and b) Rt3-OH structures. Mg blue, O red, H green.

FIG. 3. (color online) 2x2 direct methods potential map of 2x2 structure with unit cell outlined and proposed 2z2-α-OH2 structure overlaid. Mg blue, O red, H green.

FIG. 4. (color online) O-1s peaks data and fitted Gaussians for Rt3 samples annealed ex-situ with additional in-situ UHV annealing as indicated.

FIG. 5. (color online) O-1s peaks data and fitted Gaussians for 2x2 samples annealed ex-situ with additional in-situ UHV annealing as indicated. The additional feature at 526eV is a charging artifact as described in the text and should be ignored.

FIG. 6. Surface excess phase diagram for the MgO (111) system

FIG. 7. (color online) DFT relaxed 1x1H structure. Mg blue, O red, H green.

FIG. 8. (color online) DFT relaxed a) Rt3-MgH and b) Rt3-OH structures. Mg blue, O red, H green.

FIG. 9. (color online) Generic 2x2-α structure model. Mg blue, O red, surface spi-type sites 1,2,3 are green

FIG. 10. (color online) DFT relaxed 2x2-α-OH2 Structure. Mg blue, O red, H green.

FIG. 11. (color online) TPSS Surface energy vs. $H_2O$ chemical potential for several trial structures with differing surface coverage of water.

FIG. 12. (color online) TPSS surface energy per 1x1 unit cell at a water chemical potential of 0eV. Blue tie line shows the result of convex hull construction. Red arrows indicate the proposed kinetic pathway.

FIG. 13a. (color online) Diagram of proposed transition from the 1x1H (solid) → Rt3-OH (dashed) structure. 1) Transfer of hydrogen atom to adjacent oxygen site 2) Desorption of one $H_2O$ molecule per Rt3 unit cell. Mg blue, O red, H green

FIG.13b. (color online) Diagram of proposed transition from the Rt3-OH (solid) → 2x2-Oepi3 α-O3 (dashed) structure. 1) Transfer of hydrogen atom to second neighbor oxygen site 2) Desorption of one $H_2O$ molecule per 2x2 unit cell 3) Shift Oxygen atoms. Mg blue, O red, H green

**Figures**

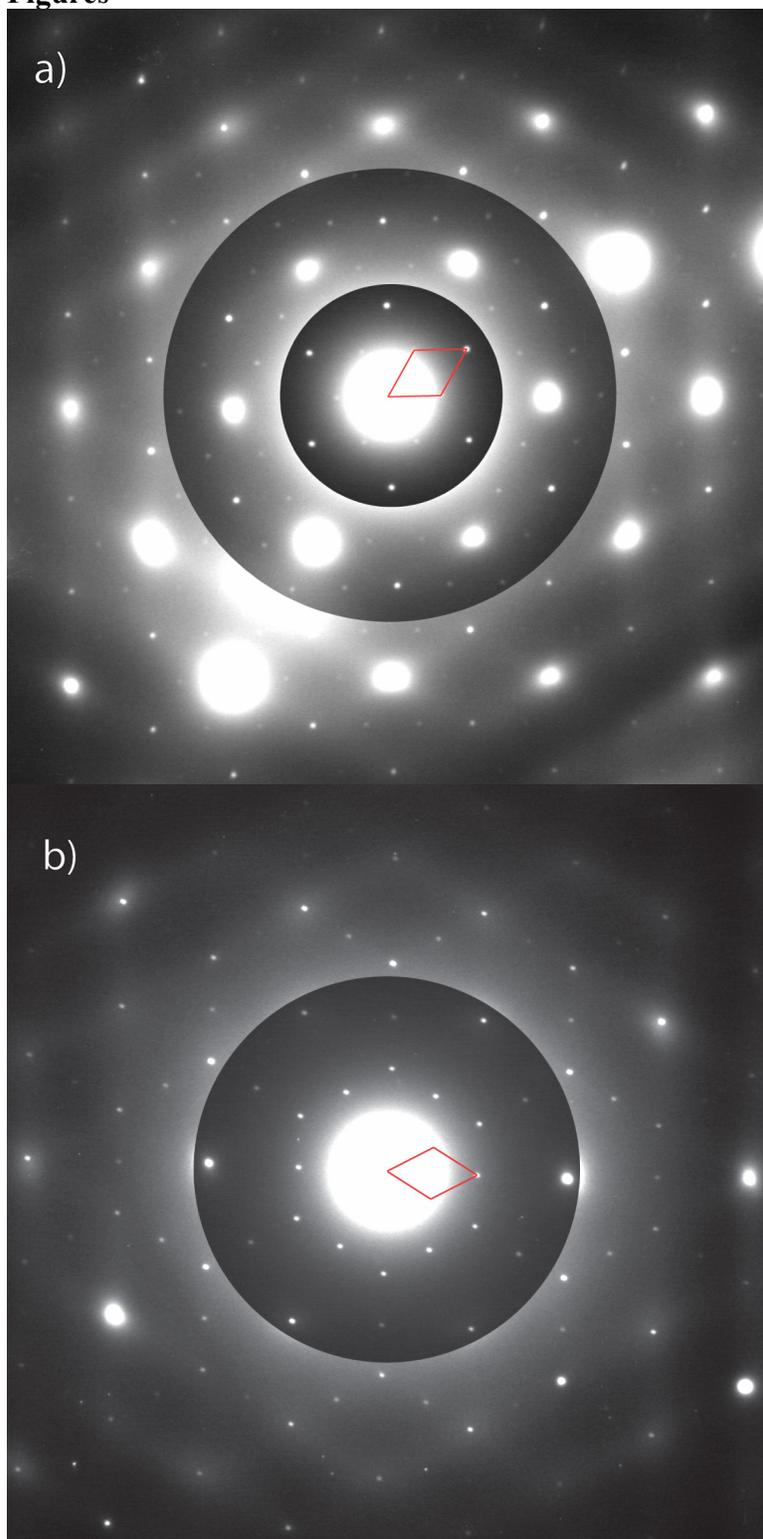

FIG. 1. Electron diffraction patterns taken from MgO samples with a) √3x√3-R30º and b) 2x2 reconstructions. The surface unit cell has been outlined for each image. Images are composites of multiple exposures to increase dynamic range.

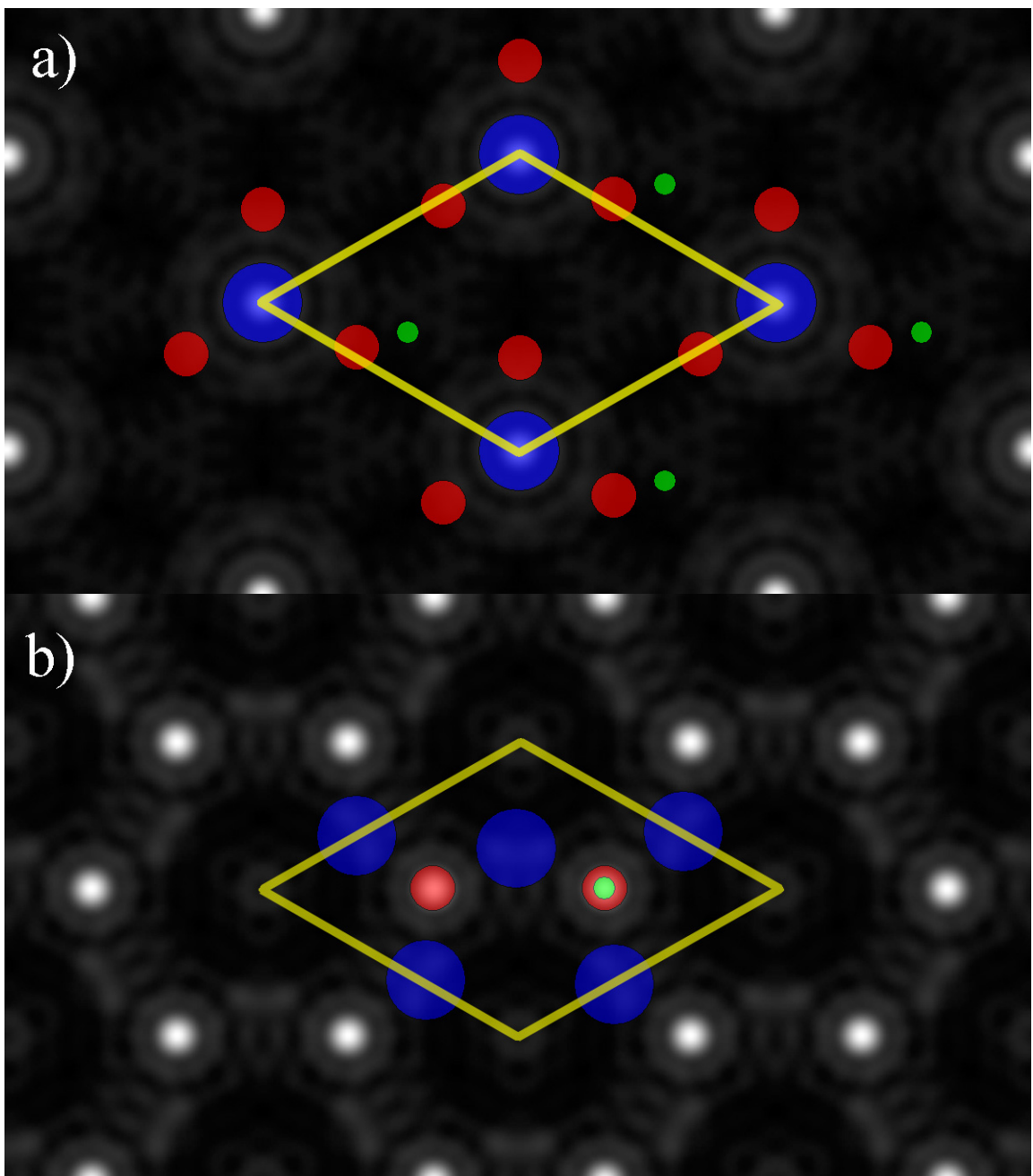

FIG. 2. (color online) Rt3 direct methods potential maps with unit cell outlined overlaid with models of the a) Rt3-MgH structure and b) Rt3-OH structures. Mg blue, O red, H green.

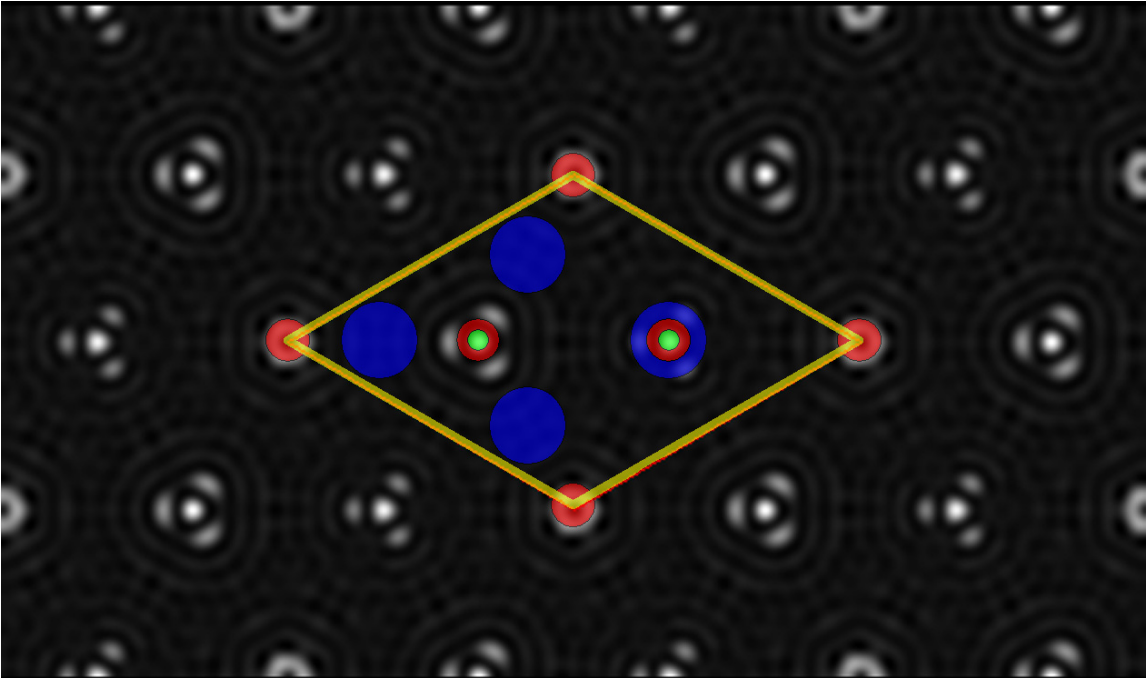

FIG. 3. (color online) 2x2 direct methods potential map of 2x2 structure with unit cell outlined and proposed 2z2-α-OH2 structure overlaid. Mg blue, O red, H green.

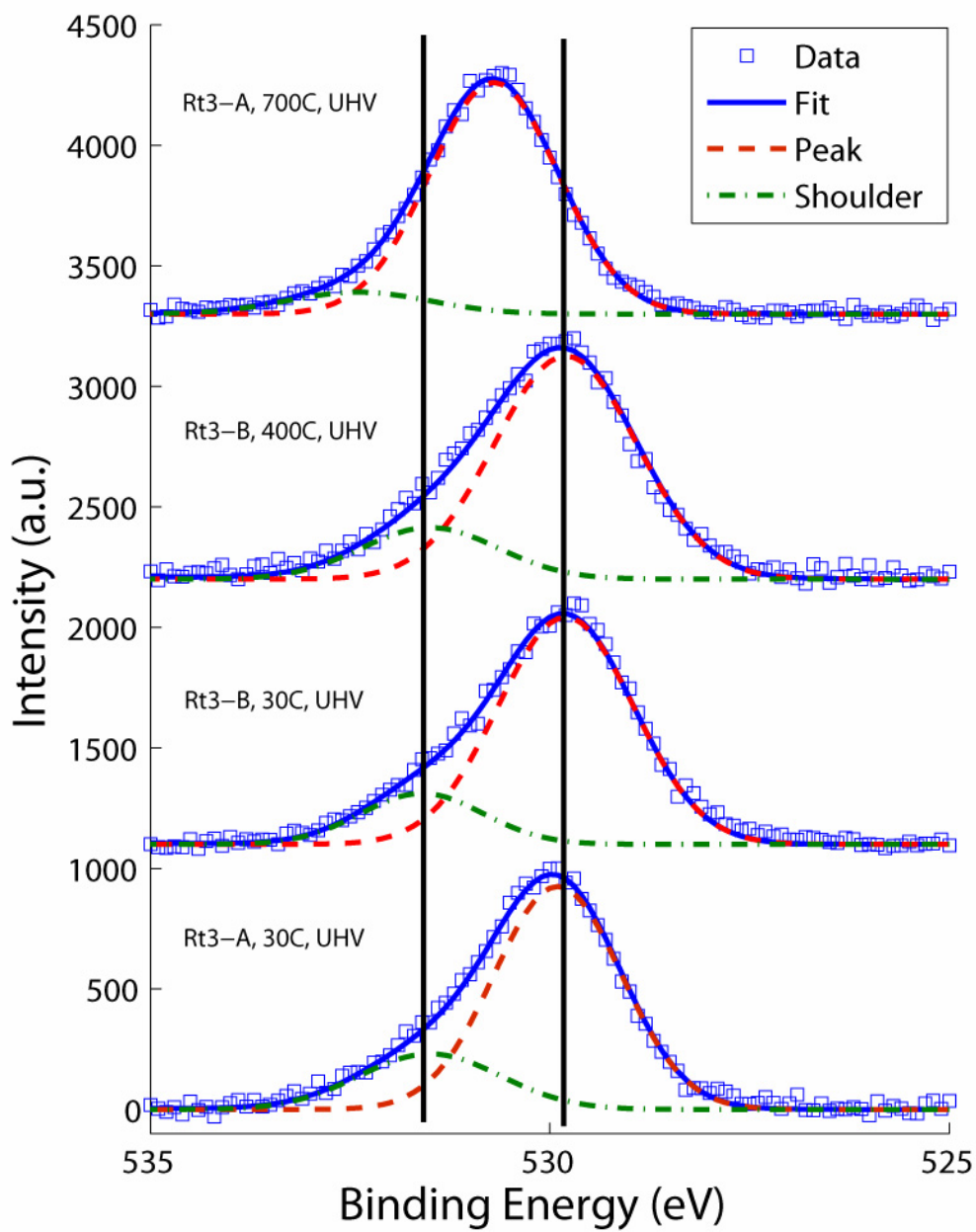

FIG. 4. (color online) O-1s data and fitted Gaussians for Rt3 samples annealed ex-situ with additional in-situ UHV annealing as indicated.

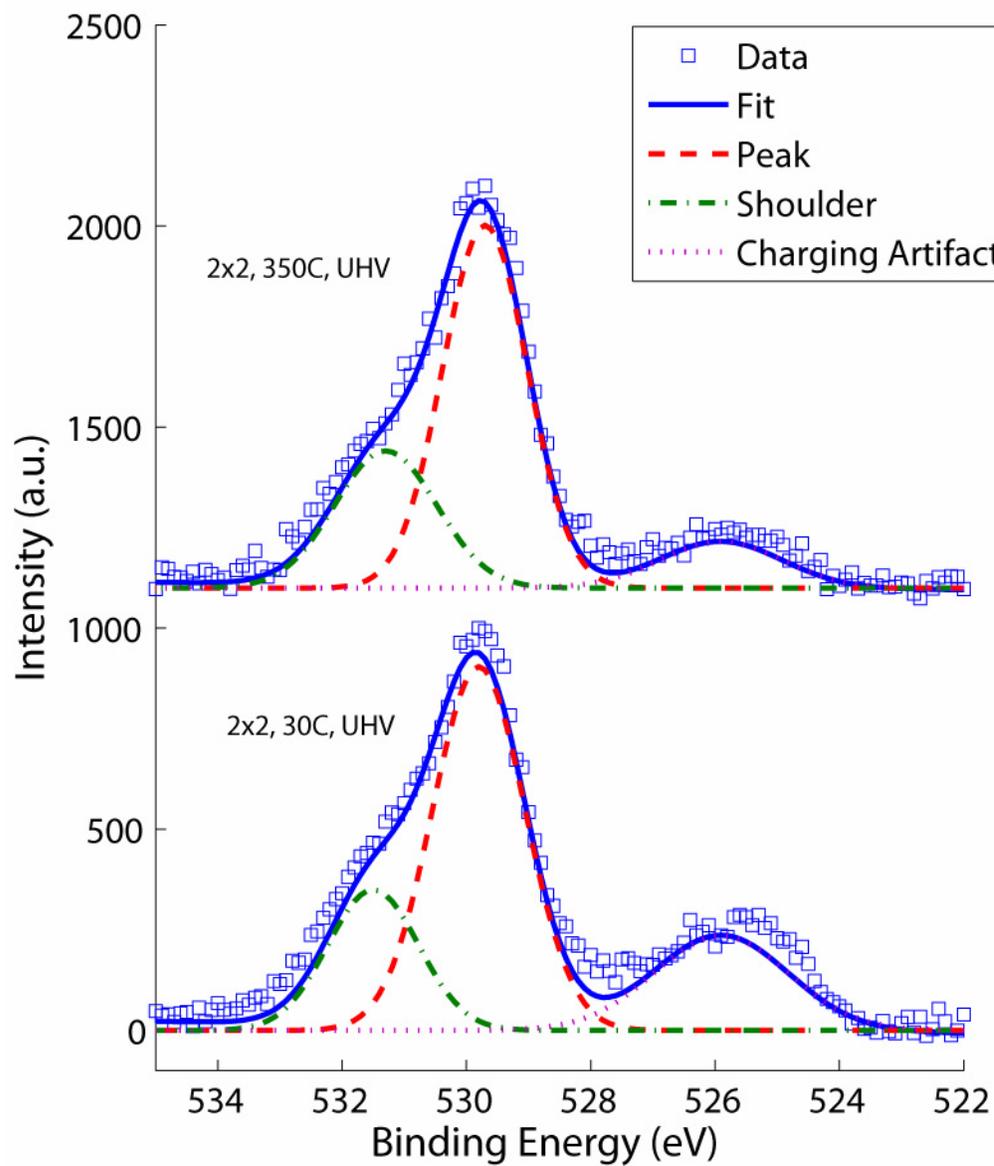

FIG. 5. (color online) O-1s data and fitted Gaussians for 2x2 samples annealed ex-situ with additional in-situ UHV annealing as indicated. The additional feature at 526eV is a charging artifact as described in the text and should be ignored.

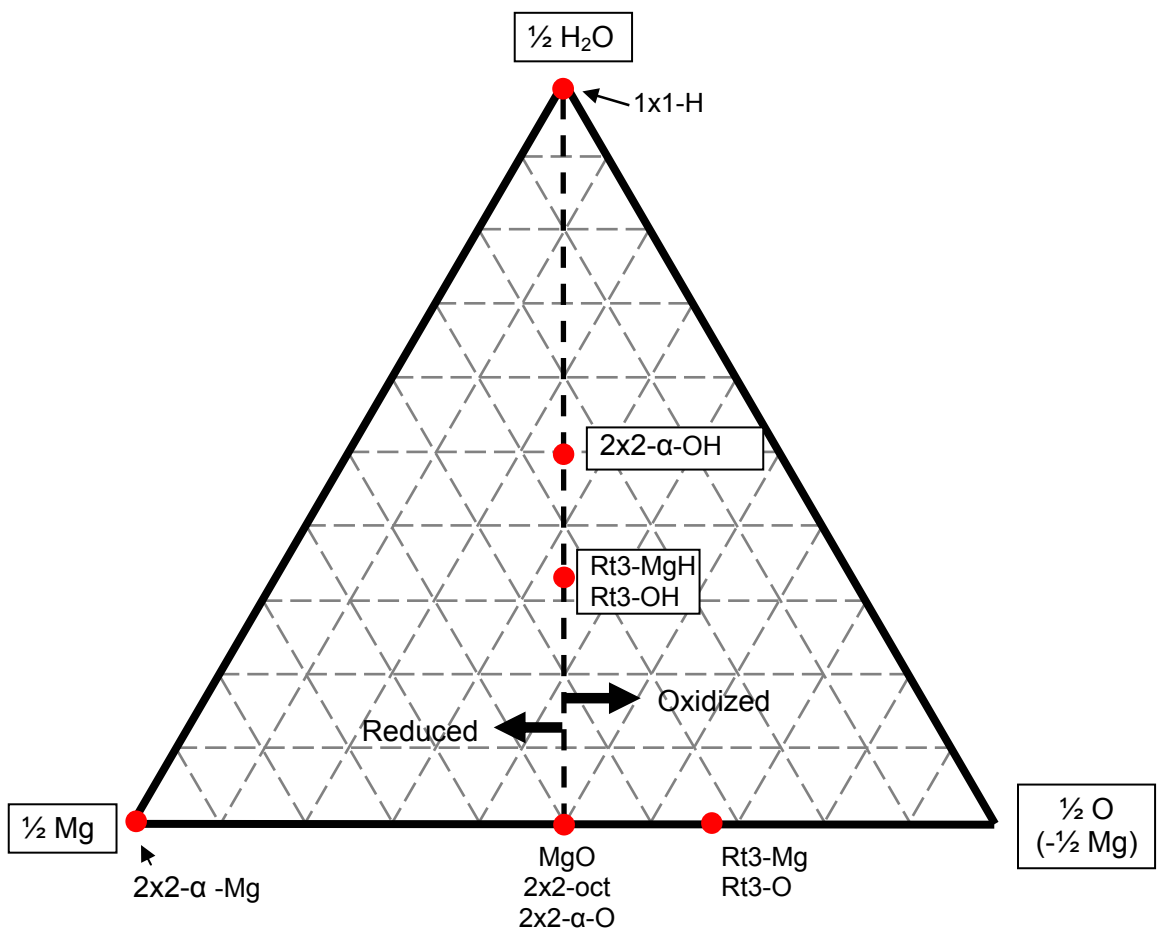

FIG. 6. Surface excess phase diagram for MgO (111) system

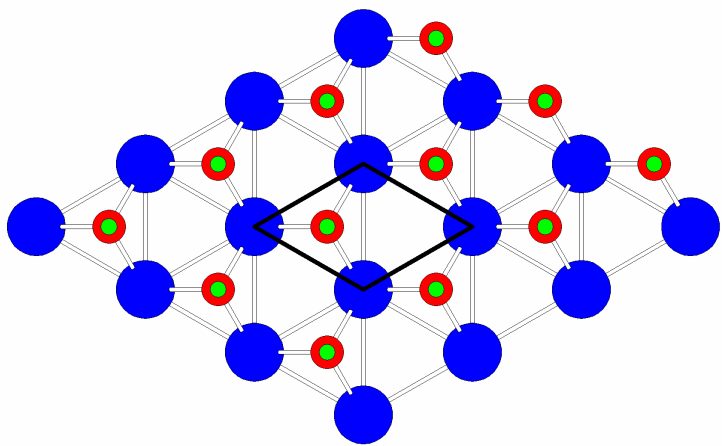

FIG. 7. (color online) DFT relaxed 1x1H structure. Mg blue, O red, H green.

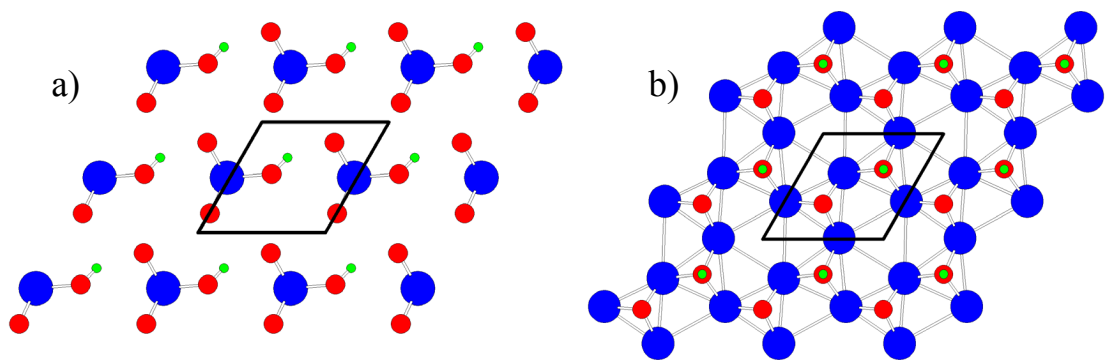

FIG. 8. (color online) DFT relaxed a) Rt3-MgH and b) Rt3-OH structures. Mg blue, O red, H green.

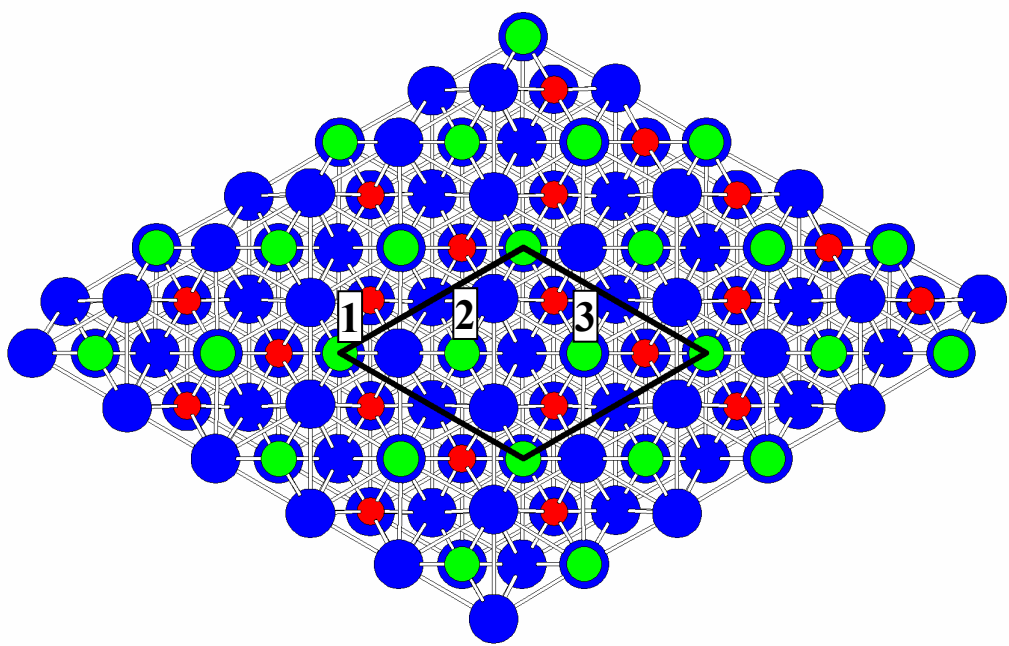

FIG. 9. (color online) Generic 2x2-α structure model. Mg blue, O red, surface sites 1,2,3 are green

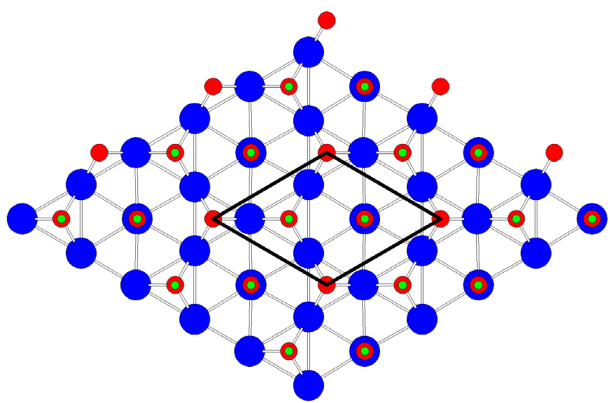

FIG. 10. (color online) DFT relaxed 2x2-α-OH2 Structure. Mg blue, O red, H green.

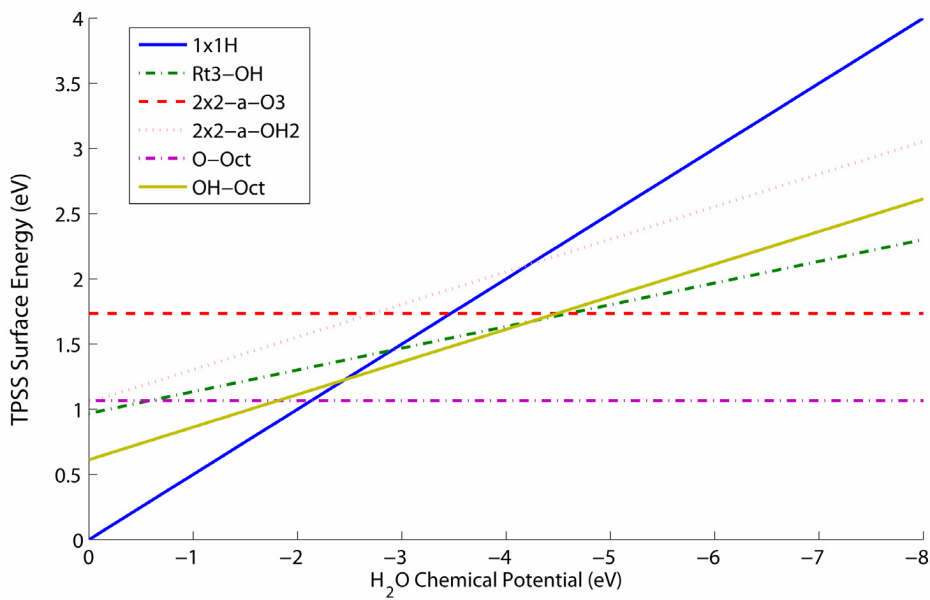

FIG. 11. (color online) TPSS Surface energy vs. $H_2O$ chemical potential for several trial structures with differing surface coverage of water.

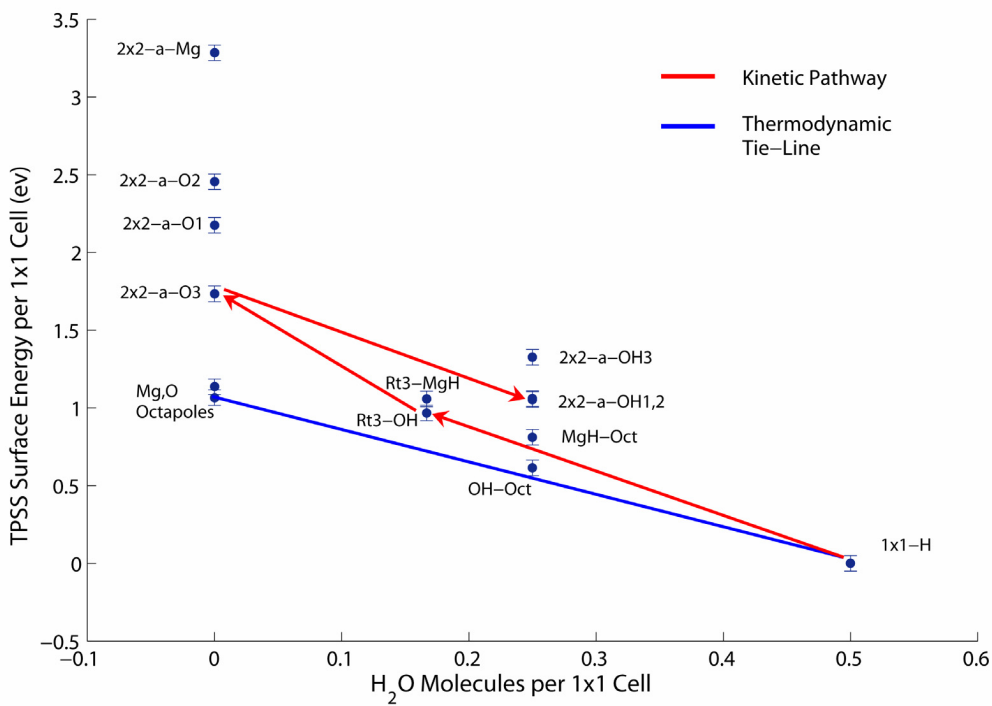

FIG. 12. (color online) TPSS surface energy per 1x1 unit cell at a water chemical potential of 0eV. Blue tie line shows the result of convex hull construction. Red arrows indicate the proposed kinetic pathway.

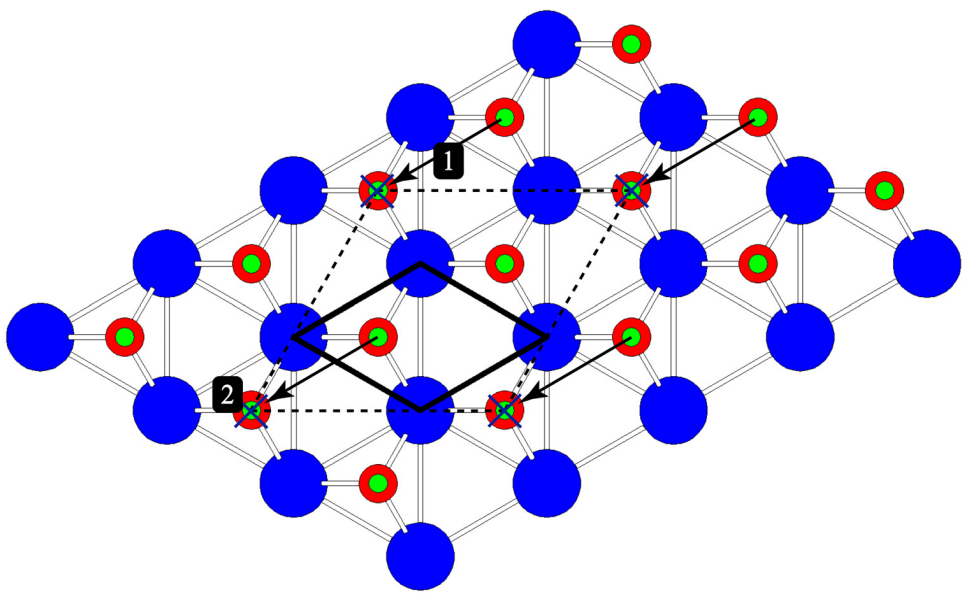

FIG. 13a. (color online) Diagram of proposed transition from the 1x1H (solid) → Rt3-OH (dashed) structure. 1) Transfer of hydrogen atom to adjacent oxygen site 2) Desorption of one $H_2O$ molecule per Rt3 unit cell. Mg blue, O red, H green

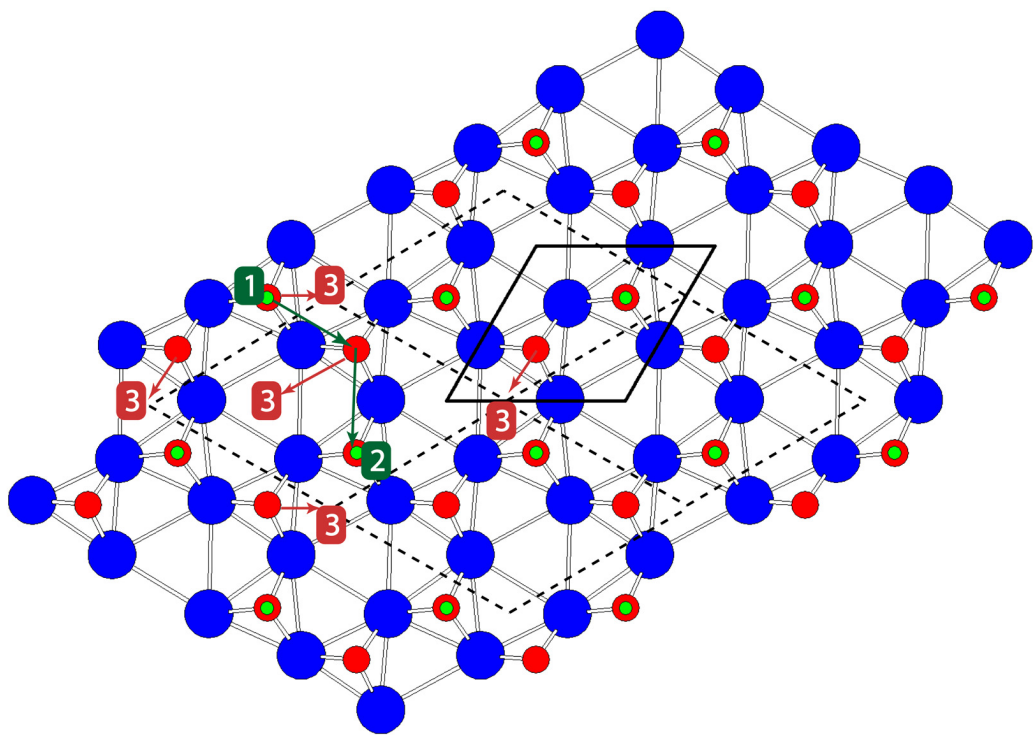

FIG. 13b. (color online) Diagram of proposed transition from the Rt3-OH (solid) → 2x2-α-O3 (dashed) structure. 1) Transfer of hydrogen atom to second neighbor oxygen site 2) Desorption of one $H_2O$ molecule per 2x2 unit cell 3) Shift Oxygen atoms. Mg blue, O red, H green